\documentclass[a4paper,11pt]{article}
\usepackage{pos}

\usepackage[export]{adjustbox}

\newcommand{\iu}{\mathrm{i}\hskip0.07em}

\title{$CPT$ and unitarity constraints for higher-order $CP$ asymmetries at finite temperature}

\author[a]{Tomáš Blažek}
\author*[a]{Peter Maták}
\author[a]{Viktor Zaujec}

\affiliation[a]{Comenius University in Bratislava,\\
 Mlynská dolina F1, 84248 Bratislava, Slovak Republic}

\emailAdd{tomas.blazek@fmph.uniba.sk}
\emailAdd{peter.matak@fmph.uniba.sk}
\emailAdd{viktor.zaujec@fmph.uniba.sk}

\abstract{
We use an unconventional diagrammatic approach to formulate $CPT$ and unitarity constraints for higher-order $CP$ asymmetries entering the source term in the Boltzmann equation. Usually, the reaction rate asymmetries in these constraints are computed within the classical kinetic theory, using zero-temperature quantum field theory to describe particles' interactions. We approximate the rates, otherwise obtained within the closed-time-path formalism, in terms of diagrams drawn on a cylindrical surface and their holomorphic cuts. The resulting equilibrium asymmetry constraints incorporate thermal-mass effects and allow tracking the cancellations of reaction rate asymmetries computed with quantum statistics. We use the top Yukawa corrections to the asymmetries in the seesaw type-I leptogenesis as an example. The contribution is primarily based on Ref. \cite{Blazek:2022azr}.
}

\FullConference{
8th Symposium on Prospects in the Physics of Discrete Symmetries (DISCRETE 2022)\\
7-11 November, 2022\\
Baden-Baden, Germany
}

\begin{document}
\maketitle

\section{Introduction}

The $CPT$ and unitarity constraints form a set of relations between reaction rate $CP$ asymmetries entering the Boltzmann equation describing particle asymmetry evolution in the early universe \cite{Dolgov:1979mz, Kolb:1979qa}. In leading-order zero-temperature calculations, these constraints come with a natural diagrammatic representation \cite{Roulet:1997xa}. However, higher-order corrections \cite{Blazek:2021olf} or, as discussed in Refs. \cite{Garbrecht:2013iga, Blazek:2021zoj}, the inclusion of thermal effects may cause substantial difficulties. In this proceedings, we briefly discuss the diagrammatic representation of the asymmetry cancellations as introduced in Ref. \cite{Blazek:2022azr} when thermal corrections are considered.

\section{{\it CP} violation at zero and finite temperature}

The violation of $CP$ requires the presence of irreducible complex phases in couplings. Furthermore, on-shell intermediate states must be present in the amplitude to produce its imaginary part represented by Cutkosky cuts over the respective Feynman diagrams. At higher orders, more on-shell cuts can be made simultaneously, and thus, for $\iu T_{ij}=S_{ij}-\delta_{ij}$ the complex conjugation in the right-hand side of the unitarity condition \cite{Kolb:1979qa}
\begin{align}\label{eq1}
\Delta\vert T_{ij}\vert^2 = \vert T_{ij}\vert^2 - \vert T_{ji}\vert^2 = 
-2\Im\Big[\big(TT^\dagger\big)_{ij}T^*{\vphantom{\big)}}_{\!\!\! ji}\Big]
+\Big\vert\big(TT^\dagger\big)_{ij}\Big\vert^2
\end{align}
cannot be omitted, spoiling the diagrammatic approach introduced in Ref. \cite{Roulet:1997xa}. This obstacle can be overcome using the holomorphic cutting rules \cite{Coster:1970jy, Bourjaily:2020wvq, Hannesdottir:2022bmo} based on an expansion of the $S$-matrix unitarity condition into a geometric series
\begin{align}\label{eq2}
(1+\iu T)^\dagger = (1+\iu T)^{-1} \quad\rightarrow\quad \iu T^\dagger = \iu T - (\iu T)^2
+ (\iu T)^3 + \ldots
\end{align}
Then, for the $CP$ asymmetry, we obtain \cite{Blazek:2021olf}
\begin{align}\label{eq3}
\Delta \vert T_{fi}\vert^2 =
&\sum_{n}(\iu T_{in} \iu T_{nf} \iu T_{fi} - \iu T_{if} \iu T_{fn} \iu T_{ni})\\
&-\sum_{n,m}(\iu T_{in} \iu T_{nm} \iu T_{mf} \iu T_{fi} - \iu T_{if} \iu T_{fm} \iu T_{mn} \iu T_{ni})+\vphantom{\sum_{n}}\ldots\nonumber
\end{align}
where the summation over the final state $\vert f\rangle$ leads to a vanishing sum of the asymmetries for a given initial state -- the $CPT$ and unitarity constraints known in the literature \cite{Dolgov:1979mz, Kolb:1979qa}. The first term in each row in Eq. \eqref{eq3} can be viewed as a forward-scattering diagram cut into several pieces. The first piece corresponds to $\iu T_{fi}$ and defines the process for which the asymmetry is calculated.

In thermal equilibrium, analogous constraints hold for asymmetries of reaction rates $\gamma_{fi}$ counting the number of $i\rightarrow f$ processes occurring per unit of volume per unit of time, which is essential to the fulfilment of Sakharov's conditions \cite{Sakharov:1967dj}. We distinguish two ways how these rates can be calculated. As a first approximation, we can use Maxwell-Boltzmann particle densities to describe the multiparticle state and zero-temperature Feynman rules for describing the interactions. In this case, we denote the rates by a small circle
\begin{align}\label{eq4}
\mathring{\gamma}_{fi}=\frac{1}{V_4}
\int\prod_{\forall i} [d\mathbf{p}_i] \mathring{f}_i(p_i) 
\int\prod_{\forall f}[d\mathbf{p}_f] 
\bigg(-\iu T_{if}\iu T_{fi} + \sum_n \iu T_{in\vphantom{f}}\iu T_{nf} \iu T_{fi}+\ldots\bigg)
\end{align}
where $[d\mathbf{p}_i] = d^3\mathbf{p}_i/((2\pi)^3 2E_i)$. Using $\mathring{f}_i(p_i)=\exp\{-E_i/T\}$, the detailed balance condition ensures
\begin{align}\label{eq5}
\sum_f\Delta\mathring{\gamma}_{fi}=0
\end{align}
inheriting the pairwise cancellations of Eq. \eqref{eq3}. 

Alternatively, we may wish to include the Bose-Einstein or Fermi-Dirac quantum phase densities and final-state statistical factors leading to uncircled rates $\gamma_{fi}$. When computing the rate asymmetries, it is impossible without modifying the cutting rules, as statistical factors also alter the on-shell parts of propagators. The correct form of the asymmetry source term can be obtained by considering the evolution of the multiparticle density matrix, as it is in the closed-time-path formalism \cite{Keldysh:1964ud, Schwinger:1960qe}. In the Markovian approximation of such evolution, the resulting rates can be expressed as an infinite series of circled rates obtained from cuttings of forward scattering diagrams drawn on a cylindrical surface \cite{Blazek:2021zoj}.

\section{Unitarity and Higgs thermal mass in right-handed neutrino decay asymmetry}

To illustrate the principle within a simple example, we consider a single forward-scattering diagram contributing to the top Yukawa corrections to the asymmetries in leptogenesis. Let us begin with the following lagrangian density
\begin{align}\label{eq6}
\mathcal{L} \supset -\frac{1}{2} M_i \bar{N}_i N_i - (Y_{\alpha i} \bar{N}_i P_L l_\alpha H +
Y_t \bar{t} P_L Q H + \mathrm{H.c.})
\end{align}
where $N_i$ corresponds to the heavy right-handed neutrino field, while $l$, $H$, $Q$, and $t$ stand for standard-model leptons, Higgs doublet, and left- and right-handed top quarks, respectively. At the $\mathcal{O}(Y^4Y^2_t)$ order, we consider a specific forward-scattering diagram cut according to Eq. \eqref{eq3}, contributing to the $NQ\rightarrow lt$ and $N_iQ\rightarrow lHQ$ circled rate asymmetries as
\begin{align}
\Delta\mathring{\gamma}_{NQ\rightarrow lt}\quad\leftarrow\quad
&\includegraphics[scale=1,valign=c]{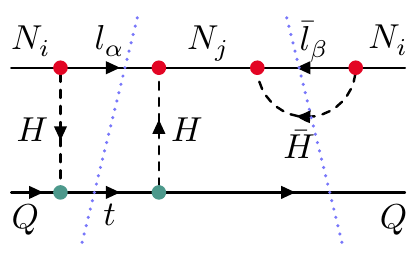} - \mathrm{m.t.}\label{eq7}\\
\Delta\mathring{\gamma}_{NQ\rightarrow lHQ}\quad\leftarrow\quad
&\includegraphics[scale=1,valign=c]{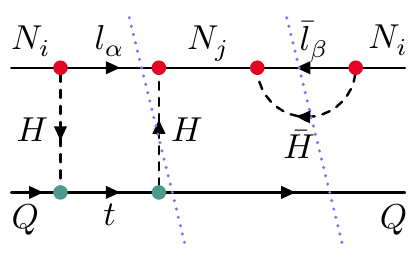}
+\includegraphics[scale=1,valign=c]{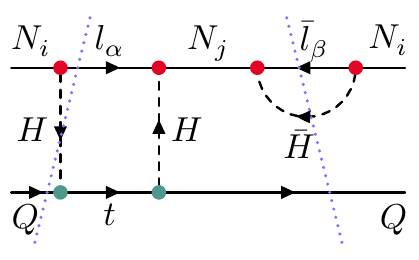}\label{eq8}\\
&-\includegraphics[scale=1,valign=c]{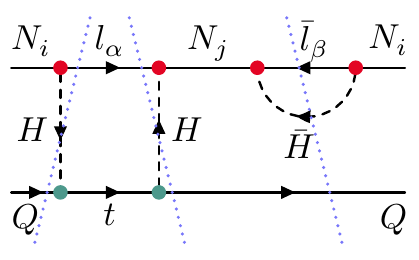} - \mathrm{m.t.}\nonumber
\end{align}
where the phase-space integration is implicitly performed as in Eq. \eqref{eq5}. The abbreviation $\mathrm{m.t.}$ stands for the \emph{mirrored terms} with the intermediate states arranged in a reversed order \cite{Blazek:2021olf}. While the calculation of the first asymmetry in Eq. \eqref{eq7} is rather straightforward, the second is slightly more subtle. When putting one of the $t$-channel Higgs propagators on its mass shell by cutting it, the second will become on-shell, too, leading to an ill-defined expression \cite{Racker:2018tzw}. To resolve this issue, we must carefully treat the intermediate states in our calculations. Using the distributional identity
\begin{align}\label{eq9}
\frac{1}{k^2+\iu\epsilon} = \mathrm{P.V.}\frac{1}{k^2}-\iu\pi\delta(k^2)
\quad\rightarrow\quad
\includegraphics[scale=1,valign=c]{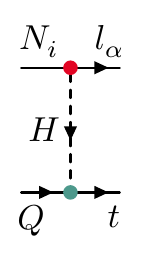} = \mathrm{P.V.}\includegraphics[scale=1,valign=c]{math6a.pdf} + \frac{1}{2}\includegraphics[scale=1,valign=c]{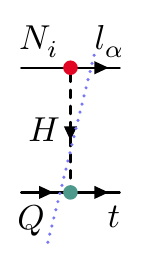}
\end{align}
we can rewrite Eq. \eqref{eq8} as
\begin{align}\label{eq10}
\Delta\mathring{\gamma}_{NQ\rightarrow lHQ}\quad\leftarrow\quad
2\mathrm{P.V.}\includegraphics[scale=1,valign=c]{math1a_cut2.pdf} - \mathrm{m.t.}
\end{align}
and apply \cite{Frye:2018xjj, Racker:2018tzw}
\begin{align}\label{eq11}
2\delta_+(k^2)\mathrm{P.V.}\frac{1}{k^2} = 
-\frac{1}{(k^0+\vert\mathbf{k}\vert)^2}\frac{\partial\delta(k^0-\vert\mathbf{k}\vert)}{\partial k^0}
\end{align}
with $k$ labelling the four-momentum of the $t$-channel Higgses. The derivative with respect to its zeroth component can be turned into the derivative with respect to the Higgs mass through \cite{Blazek:2021gmw}
\begin{align}\label{eq12}
\frac{\partial}{\partial k^0}\bigg\vert_{k^0=\vert\mathbf{k}\vert} \frac{\mathcal{F}(k^0,\mathbf{k})}{(k^0+\vert\mathbf{k}\vert)^2}=
\frac{\partial}{\partial m^2_H} \bigg\vert_{m^2_H=0}\frac{\mathcal{F}(E_{\mathbf{k}},\mathbf{k})}{2E_{\mathbf{k}}}
\quad\text{for}\quad E_{\mathbf{k}}=\sqrt{m^2_H+\mathbf{k}^2}
\end{align}
such that
\begin{align}\label{eq13}
\Delta\mathring{\gamma}_{N_iQ\rightarrow lHQ\vphantom{)}}=
\Delta\mathring{\gamma}_{N_i(Q)\rightarrow lH(Q)}+
\frac{1}{4}\mathring{m}^2_{H,Y_t}(T)\frac{\partial}{\partial m^2_H} \bigg\vert_{m^2_H=0}\Delta\mathring{\gamma}_{N_i\rightarrow lH}
\end{align}
where the Higgs thermal mass
\begin{align}\label{eq14}
\mathring{m}^2_{H,Y_t}(T)=12 Y^2_t  \int [d\mathbf{p}_Q] \mathring{f}_Q
\end{align}
is obtained from classical or circled phase-space density. In Eq. \eqref{eq13} we denote
\begin{align}\label{eq15}
\Delta\mathring{\gamma}_{N_i(Q)\rightarrow lH(Q)}= 
\includegraphics[scale=1,valign=c]{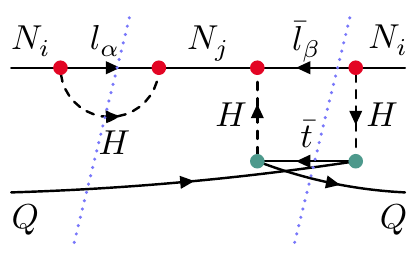}
+\includegraphics[scale=1,valign=c]{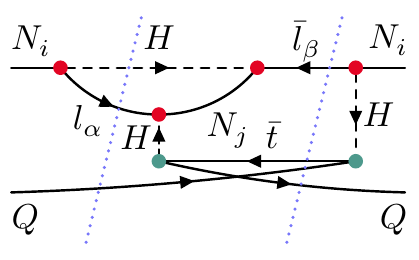}
-\hskip2mm\mathrm{m.t.}
\end{align}
approximating the Pauli blocking factor of $Q$ in the asymmetry-generating loop contributing to the $N_i\rightarrow lH$ asymmetry. This can be observed in
\begin{align}\label{eq16}
\includegraphics[scale=1,valign=c]{math1c_cut1.pdf} = -\mathring{f}_Q \times\includegraphics[scale=1,valign=c]{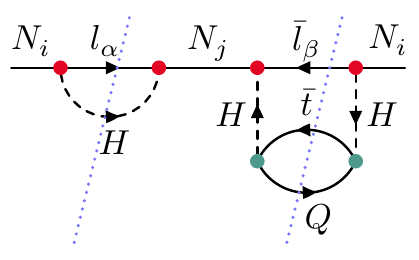}
\end{align}
before the initial-state momentum integration is carried out. Finally, considering the full list of forward-scattering diagrams of the same perturbative order allows us to write the $CPT$ and unitarity constraints \cite{Blazek:2022azr, Blazek:2021olf}
\begin{align}\label{eq17}
\Delta\mathring{\gamma}_{N_iQ\rightarrow lt\vphantom{\bar{l}}}+
\Delta\mathring{\gamma}_{N_i(Q)\rightarrow lH(Q)\vphantom{\bar{l}}}+
\Delta\mathring{\gamma}_{N_i(Q)\rightarrow \bar{l}\bar{H}(Q)}+
\Delta\mathring{\gamma}_{N_iQ\rightarrow \bar{l}QQ\bar{t}} = 0
\end{align}
and a separately vanishing mass-derivative of the right-handed neutrino decay asymmetries
\begin{align}\label{eq18}
\frac{1}{4}\mathring{m}^2_{H,Y_t}(T)\frac{\partial}{\partial m^2_H} \bigg\vert_{m^2_H=0}\bigg(\Delta\mathring{\gamma}_{N_i\rightarrow lH\vphantom{\bar{H}}} + \Delta\mathring{\gamma}_{N_i\rightarrow\bar{l}\bar{H}}\bigg)=0.
\end{align}
It is remarkable that even though starting with classical kinetic theory and zero-temperature Feynman rules, an approximation of thermal mass effect in the decay asymmetry kinematics inconspicuously entered the computation. We emphasize it is a natural consequence of unitarity and the so-called anomalous thresholds \cite{Hannesdottir:2022bmo, Karplus:1958zz, Karplus:1959zz, Nambu:1958zze}. In our calculation, they entered Eq. \eqref{eq8} and were represented by cuts dividing the amplitude into connected and disconnected diagrams.

We may wish to generalize the relations in Eqs. \eqref{eq17} and \eqref{eq18} to include quantum densities as well as uncircled thermal mass. For that purpose, we consider a diagram similar to that in Eq. \eqref{eq7} drawn on a cylindrical surface \cite{Blazek:2022azr}
\begin{align}\label{eq19}
\includegraphics[scale=1,valign=c]{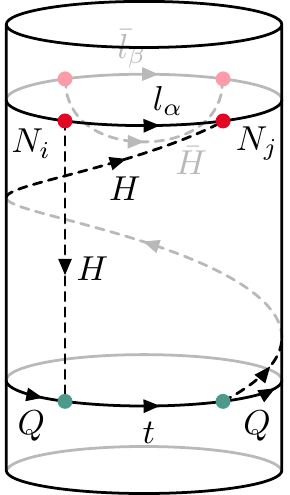} \quad\rightarrow\quad
\includegraphics[scale=1,valign=c]{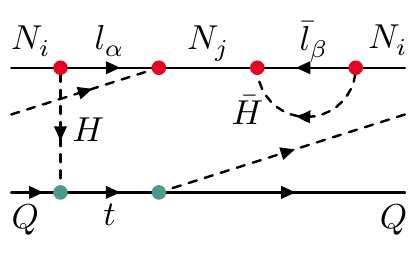}
\end{align}
where the winding number of the Higgs internal line has been increased to one. Then, the resulting forward-scattering diagram is cut in all possible ways according to Eq. \eqref{eq3} and a set of relations analogous to Eqs. \eqref{eq17} and \eqref{eq18} is obtained. Summation of resulting circled asymmetries obtained from cylindrical diagrams with all possible winding numbers of all lines leads to a simple replacement
\begin{align}\label{eq20}
\mathring{f}_{N_i\vphantom{\bar{H}}} \mathring{f}_{Q\vphantom{\bar{H}}} 
\quad\rightarrow\quad
f_{N_i\vphantom{\bar{H}}}f_{Q\vphantom{\bar{H}}} (1+f_{H\vphantom{\bar{H}}})(1-f_{l\vphantom{\bar{H}}}) (1+f_{\bar{H}})(1-f_{\bar{l}})
\end{align}
to be made in Eq. \eqref{eq7}, leading to uncircled asymmetry of the  $NQ\rightarrow lt$ reaction \cite{Blazek:2022azr}. Finally, applying the same procedure to all diagrams contributing to the asymmetry at the given perturbative order, we may erase the circles in Eqs. \eqref{eq17} and \eqref{eq18}, and thus, the $CPT$ and unitarity constraints for the $CP$ asymmetries including thermal corrections have been obtained.

\section{Conclusions}

In this contribution, a new diagrammatic method \cite{Blazek:2021zoj} has been applied to the example of the $\mathcal{O}(Y^4Y^2_t)$ asymmetries of the seesaw type-I leptogenesis with right-handed neutrino and left-handed third-generation quark in the initial state. The systematic procedure leading to the $CPT$ and unitarity constraints for such asymmetries, in which thermal effects are included, has been described briefly. The details, including explicit expressions for the $CP$ asymmetries, can be found in Ref. \cite{Blazek:2022azr}.

\section*{Acknowledgements}
The authors have been supported by the Slovak Ministry of Education Contract No. 0466/2022 and by the Slovak Grant Agency VEGA, project No. 1/0719/23. Viktor Zaujec received funding from Comenius University in Bratislava, project No. UK/317/2022: \emph{Unitarita, CP asymetrie a diagramatick\'{a} reprezent\'{a}cia tepeln\'{y}ch efektov v leptogen\'{e}ze}. We thank our colleague Fedor \v{S}imkovic for his long-term support.

\bibliographystyle{JHEP}
\bibliography{peter_matak.bib}

\end{document}